\documentclass[conference]{IEEEtran}
\IEEEoverridecommandlockouts
\usepackage{subcaption}
\usepackage{cite}
\usepackage{amsmath,amssymb,amsfonts}
\usepackage{algorithmic}
\usepackage{graphicx}
\usepackage{textcomp}
\usepackage{xcolor}
\def\BibTeX{{\rm B\kern-.05em{\sc i\kern-.025em b}\kern-.08em
    T\kern-.1667em\lower.7ex\hbox{E}\kern-.125emX}}
\usepackage{pdfpages}
\usepackage{algorithmic}
\usepackage{tabularx}
\usepackage{graphicx}
\usepackage{textcomp}
\usepackage{siunitx}
\usepackage{xcolor}
\usepackage{csquotes}
\usepackage[hyphens]{url}
\usepackage{cleveref}
\usepackage{tikz}
\usepackage{pgfplots}
\usepackage{truncate}
\usepackage{collcell}
\usepackage{booktabs}
\usepackage{float}
\usepackage{tabularx}
\newcommand*\truncatecell[1]{\truncate{5.5cm}{#1}}
\newcolumntype{Y}{>{\collectcell\truncatecell}p{5.5cm}<{\endcollectcell}}
\usetikzlibrary{shapes.geometric,decorations.pathmorphing}
\usetikzlibrary{positioning, arrows.meta}
\def\BibTeX{{\rm B\kern-.05em{\sc i\kern-.025em b}\kern-.08em
    T\kern-.1667em\lower.7ex\hbox{E}\kern-.125emX}}

\pgfplotsset{every axis/.append style={font=\sffamily}}
\pgfplotsset{every axis label/.append style={font=\sffamily}}
\pgfplotsset{every tick label/.append style={font=\sffamily}}
\tikzset{every picture/.style={/utils/exec=\sffamily}}

\newcommand*\emptycirc[1][1ex]{\tikz\draw (0,0) circle (#1);} 

\newcommand*\fullcirc[1][1ex]{\tikz\fill (0,0) circle (#1);} 

\begin{document}

\title{SwarmSearch: Decentralized Search Engine \\with Self-Funding Economy
\thanks{
This work has been submitted for possible publication. Copyright may be transferred without
notice, after which this version may no longer be accessible.
This work was funded by the Dutch National NWO/TKI Science Grant BLOCK.2019.004.}
}
\author{
\IEEEauthorblockN{
Marcel Gregoriadis, Rowdy M Chotkan, Petru Neague, and Johan Pouwelse
}
\IEEEauthorblockA{
Delft University of Technology, Delft, The Netherlands\\
\{m.gregoriadis, r.m.chotkan-1, p.m.neague, j.a.pouwelse\}@tudelft.nl
}
}

\maketitle

\begin{abstract}

Centralized search engines control what we see, read, believe, and vote.
Consequently, they raise concerns over information control, censorship, and bias.
Decentralized search engines offer a remedy to this problem,
but their adoption has been hindered by their inferior quality and lack of a self-sustaining economic framework.
We present SwarmSearch, a fully decentralized, AI-powered search engine with a self-funding architecture.
Our system is designed for deployment within the decentralized file-sharing software Tribler.
SwarmSearch integrates volunteer-based with profit-driven mechanisms to foster an implicit marketplace for resources.
Employing the state-of-the-art of AI-based retrieval and relevance ranking, we also aim to close the quality gap between decentralized search and centralized alternatives.
Our system demonstrates high retrieval accuracy while showing robustness in the presence of 50\,\% adversarial nodes.
\end{abstract}


\section{Introduction}

Decentralized search engines offer a remedy against the risks posed by Big Tech's centralized control over information. 
The dominance of platforms like Google raise concerns around censorship, privacy violations, and bias.
Recently, the U.S. court ruled that Google is a monopoly~\cite{usdoj_google_2023}.
This position enabled Google to shape access to information, introducing censorship and algorithmic bias to search results~\cite{epstein2016new,solon2016google,stjernfelt2020facebook}.
Other Big Tech firms faced similar scrutiny. 
Meta's Cambridge Analytica scandal revealed mass misuse of user data for political targeting~\cite{hinds2020wouldn,stjernfelt2020facebook}.
TikTok is currently under investigation by the European Commission for alleged election interference via its recommender system~\cite{eu2024tiktok}.

Decentralized search resolves the issues around centralized control.
In a system for decentralized search, documents are distributed across a network of independent nodes (or \emph{agents}).
Documents, in this context, may represent metadata on entities such as torrents or websites.
Each agent serves as an index over the documents it knows about.
At the same time, every agent can issue queries to the network to find relevant documents it does \emph{not} know about.
Agents collaborate on tasks such as routing and retrieval,
facilitating a global search network that scales and coordinates autonomously without centralized control.
However, while decentralized file-sharing systems like BitTorrent and IPFS are flourishing, research on decentralized search has mostly stalled.
Despite decades of research on decentralized search~\cite{chawathe2003making,keizer2024survey,khudhur2019siva,li2021bringing,loo2005case,wang2020keyword,lu2002panache},
a fully decentralized and viable search engine has remained elusive.
It has proven extremely challenging to design a decentralized search engine capable of matching the quality of centralized systems while also having the financial architecture to sustain its service.

Existing decentralized search engines like YaCy~\cite{herrmann2014description} are nowhere near the quality of centralized search engines, as the gap between their outdated information retrieval techniques and modern AI-driven algorithms keeps widening~\cite{onal2018neural}. 
Many decentralized initiatives compromise their vision by reintroducing centralized elements to simplify technical hurdles~\cite{archiveBumpRoad,poort2014baywatch,presearchPresearch,li2021bringing}.
IPFS-Search, for instance, depended on centralized servers for crawling and indexing, but collapsed when funding ran dry~\cite{archiveBumpRoad}.
While centralized search engines sustain their operation through data-driven business models,
decentralized systems face the challenge of having to motivate users to perform operations on behalf of others.
Efforts to motivate user contributions have often relied on altruism or socio-psychological rewards~\cite{wei2015motivating,goes2016incentive}.
When motivation is made extrinsic, e.g., through crypto-economic incentives, this encourages low-quality contributions and spam~\cite{qiao2021mitigating}.
Furthermore, users demonstrate a strong preference for zero-cost options, which has led to the failure of micropayment models in the past~\cite{ecbMicropayments,solaiman2024internet}.
The core challenge lies in designing a decentralized financial architecture that recognizes and rewards high-quality contributions, but refrains from imposing mandatory fees on its users.

\begin{table*}[ht]
\centering
    \vspace{0.03in}
  \caption{Characteristics of Decentralized Search Engine Initiatives}
  \label{tab:freq}
    \begin{tabular}{lccccccc}
    \toprule
Search Engine & Deployment & Self-funded & Free-to-use & Spam-resilient & Self-learning & Semantic Search & No central comp. \\
\midrule
YaCy (2004)~\cite{herrmann2014description}     & \fullcirc & \emptycirc & \fullcirc & \emptycirc & \emptycirc & \emptycirc & \fullcirc \\
MAAY (2006)~\cite{ngoc2006maay}               & \emptycirc & \emptycirc & \fullcirc & \emptycirc & \fullcirc & \emptycirc & \fullcirc \\
Tribler (2006)~\cite{pouwelse2008tribler}     & \fullcirc & \emptycirc & \fullcirc & \emptycirc & \emptycirc & \emptycirc & \fullcirc \\
PreSearch (2017)~\cite{presearchPresearch}    & \fullcirc & \fullcirc & \fullcirc & \fullcirc & \emptycirc & \emptycirc & \emptycirc \\
The Graph (2018)~\cite{thegraph}              & \fullcirc & \fullcirc & \emptycirc & \fullcirc & \emptycirc & \emptycirc & \fullcirc \\
DeSearch (2021)~\cite{li2021bringing}         & \emptycirc & \fullcirc & \emptycirc & \emptycirc & \emptycirc & \emptycirc & \emptycirc \\
\midrule
\textsc{SwarmSearch} (2025) & \fullcirc & \fullcirc & \fullcirc & \fullcirc & \fullcirc & \fullcirc & \fullcirc \\
\bottomrule
\end{tabular}
\label{tab:searchEngines}
\end{table*}

We introduce \textsc{SwarmSearch}, a self-funding and fully decentralized search engine powered by a decentralized network of AI agents that collaboratively perform search tasks and learn from each other.
Using state-of-the-art models for document retrieval and relevance ranking, we aim to close the quality gap with centralized systems.
Our self-funding mechanism, inspired by Bitcoin’s mining incentives and volunteer-driven platforms like Linux and Wikipedia, uses a donation-based system 
to ensure viability without mandatory fees.
By combining profit-driven incentives with volunteer-based contributions, we create a financial architecture that refrains from mandatory usage fees, yet facilitates the resource sharing critical for delivering high-quality search results.
Our financial architecture, therefore, emerges as an \emph{implicit} marketplace for resources.

This work is conducted in the context of a 20-year running research line on Internet-deployed decentralized systems. With our \emph{systems lab} background, we aim to bridge the gap between theoretical designs and systems used by billions.
With \textsc{SwarmSearch}, we present the proto-design for deployment within the Tribler peer-to-peer file-sharing application~\cite{pouwelse2008tribler}.
As such, we are committed to building a production-ready system.
We hope to leverage the AI revival of peer-to-peer systems~\cite{gregoriadis2025decentralized,neague2024dsi,neague2025semantica} to foster a cooperative global search network, finally offering a competitive and viable alternative to centralized systems.

The remainder of this paper is organized as follows.
\Cref{sec:problem} formulates the problem. 
\Cref{sec:related} describes related work.
\Cref{sec:system} specifies our system model.
In \Cref{sec:design}, we detail the design of our search engine.
\Cref{sec:dataset} introduces a dataset of real-world search logs, which we use in our experiments.
In \Cref{sec:exp}, we evaluate our system with regards to retrieval accuracy and robustness against attacks.
We conclude in \Cref{sec:conclusion}.

\section{Problem Description}\label{sec:problem}

A decentralized search system with self-funding has never been achieved.
Numerous distributed systems with keyword search have been launched in the past 25 years.
The first generation of peer-to-peer systems used exact syntax matching of filenames, for instance, Gnutella~\cite{ripeanu2001peer}.
The fully self-organising approach of Gnutella with search broadcasting proved to be unsustainable.
These early systems lacked any mechanism against freeriding~\cite{adar2000free}.
There has been a consistent failure in devising self-funded architectures. 
For instance, the fully decentralized storage system IPFS lacks search.
Central servers are required to crawl this network and provide search. 
Yet, no business model was found for self-funding of these servers.
IPFS-Search was forced to shut down after running out of initial funding~\cite{archiveBumpRoad}.

A comparison with state-of-the-art centralized search engines such as Google and Bing highlights the widening gap in quality between centralized and decentralized search.
This gap is driven by the ability of centralized systems to aggregate massive amounts of user data, employ learning algorithms, and leverage semantic matching to interpret and satisfy user intent.
Market leadership in targeted advertisement has lifted Google to a USD 2 trillion market capitalization~\cite{Alphabet2025SEC}.
Such financial resources allow Google and other centralized search engines to operate the costly data centers that underpin their search capabilities.

Bitcoin's double-spending prevention is funded by Bitcoin mining.
The breakthrough of Bitcoin was the complete decentralization of a circular financial ecosystem. It proved that money could exist without requiring banks, central banks, or even governments. Bitcoin mining is the first example of a self-funded network.
The self-funding mechanisms pioneered by the blockchain community are used for funding network services (e.g., double-spending prevention).
This approach cannot easily be generalized to search.
Users are unwilling to pay for digital services, as in the form of micropayments, when alternatives exist for free~\cite{ecbMicropayments,solaiman2024internet}.
The Freemium business model was suggested informally as far back as 2008 to
provide self-funding for peer-to-peer services~\cite{p2pfoundationFreemiumBusiness}.
The Freemium model offers basic services free of charge, while more advanced features must be paid for~\cite{rietveld2018creating}.
Unfortunately, this promising self-funding model has never been realized in real systems.

We conclude that the problem for decentralized search lies in the absence of a system that satisfies the following properties:
\begin{itemize}
    \item It must be \emph{self-funded} in order to facilitate high-quality contributions, whilst also being \emph{spam-resilient}.
    \item At the same time, the service must be \emph{free-to-use}.
    \item In order to compete with centralized alternatives, it must be capable of \emph{semantic search} and incorporate \emph{self-learning} behavior in order to adapt to user contexts.
\end{itemize}

\begin{figure*}[t]
    \centering
    \includegraphics[width=\linewidth]{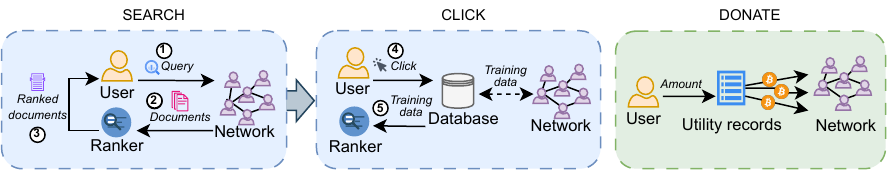}
    \caption{Overview of core user actions and architecture of SwarmSearch.}
    \label{fig:search}
\end{figure*}

\section{Related Work}\label{sec:related}

We find that most research on decentralized search is rooted in the peer-to-peer (P2P) era of the early 2000s~\cite{chawathe2003making,loo2005case,gassend2001dinx}, after which interest in the field has notably waned.
The Web3 movement of recent years has caused a little revival of the field~\cite{wang2020keyword,khudhur2019siva,li2021bringing}.
A recent survey by Keizer et al.~\cite{keizer2024survey} revealed, however, that no current system adequately addresses these issues in a comprehensive manner.
They conclude that even though numerous projects for decentralized search have been proposed, they generally focus on narrow aspects of the problem.
Consequently, in practice, users still rely on centralized indices to locate files within decentralized storage networks, as exemplified by IPFS-Search~\cite{archiveBumpRoad}.

Some researchers have proposed to decentralize IPFS-Search by maintaining the extracted metadata on the DHT~\cite{khudhur2019siva,zhu2020keyword,wang2020keyword,keizer2023ditto}.
The initial retrieval and ranking are typically performed based on exact keyword matching and term-based metrics; hence, they rely on exact term matches.
Other researchers proposed distributing content based on locality-sensitive hashing (LSH), thereby supporting syntactic similarity search~\cite{fujita2021similarity,keizer2023ditto}, but not \emph{semantic} similarity search.
To the best of our knowledge, De-DSI~\cite{neague2024dsi} is the first work to leverage language models to enable semantic search.
Recently, DART~\cite{gregoriadis2025decentralized} proposed decentralized relevance ranking based on deep learning.
Feedback-driven algorithms like De-DSI and DART allow the search engine to adapt to user preferences and opinion drift. 
MAAY~\cite{ngoc2006maay} and G-Rank~\cite{gold2023g} employ fixed heuristics to aggregate user data for collaborative filtering.

Research on decentralized search has so far ignored the problem of motivating user contributions, typically assuming both honest and altruistic peers.
YaCy~\cite{herrmann2014description} and the search in Tribler~\cite{pouwelse2008tribler} are examples of deployed decentralized search engines that rely on these assumptions.
The Graph~\cite{thegraph} is a decentralized search engine specifically for data on the blockchain.
While The Graph is spam-resistant, its financial architecture requires users to pay for each query they submit.
PreSearch~\cite{presearchPresearch}, on the other hand, is a project that is both self-funded and free-to-use.
It rewards both usage and service, and is funded by advertising expenditures.
However, PreSearch relies on a central gateway server for functions like user anonymization and reputation~\cite{presearchWhatNode}.

In \Cref{tab:searchEngines}, we show a list of decentralized search engine initiatives, both from industry and academia, and their characteristics.
Specifically, we show how they satisfy the properties outlined in \Cref{sec:problem}.

\section{System Model}\label{sec:system}

Our self-funded search system mixes elements from volunteer-based computing with profit-driven mining known from the Bitcoin world.
By doing so, we aim to replicate the emergent effect from initiatives such as Linux, Wikipedia, BitTorrent, Kickstarter, and Bitcoin.
In this section, we present our system model.
The terms \emph{user}, \emph{agent}, and \emph{node} are used interchangeably throughout this paper. While we use \emph{node} to denote a network entity, \emph{user} for an initiator of an action, and \emph{agent} for a responder, these distinctions are purely contextual--there are no fixed roles in the system.

Our system considers a network of nodes, which form a connected graph. 
Assuming the graph remains connected, nodes may join or leave the network at any time. 
We assume nodes engage in decentralized searches for websites,
i.e., searching relevant URLs for a given query.
We illustrate the system model through three core user actions (see~\Cref{fig:search}):
\begin{enumerate}
    \item \textbf{Search:} A user sends a query to a sample of agents, each of which responds with a set of matching documents. Locally, the documents are aggregated and ranked before being presented to the user.
    \item \textbf{Click:} After being presented with a list of search results, the user performs a \emph{click} on one of the results. Clicklogs are collected in a local database and asynchronously exchanged with other peers.
    \item \textbf{Donate:} At any point, a user can donate an arbitrary amount of Bitcoin. The amount is distributed among peers in proportion to the \emph{utility} of their past contributions to the user.
\end{enumerate}

Our system pays agents for providing useful work.
Instead of imposing usage fees, however, rewards are funded through donations made by altruistic users.
Users in our system provide \emph{utility} to others via two activities: document retrieval and exchange of training data (in the form of \emph{clicklogs}).
Each user maintains \emph{utility scores} that reflect the utility received from other peers.
Donations (in the form of Bitcoin) are distributed based on locally maintained utility records.
This creates an implicit marketplace for resources.
Users who actively donate to the system will attract more agents who are willing to provide utility to them, even when they do not contribute resources themselves.
We expect this to lead to the emergence of high-tier and low-tier service levels that reflect the user's conversion rate with the system.
Users who do not choose to donate can use the search engine for free while monetizing their resources.
Importantly, the system remains viable even if only a small fraction chooses to donate.

\section{Design of SwarmSearch}\label{sec:design}

We introduce \textsc{SwarmSearch}, the first self-funded decentralized search engine.
Our search engine is powered by a self-learning network of autonomous AI agents that collaboratively perform search tasks and exchange training data.
With our novel financial architecture, we create an implicit marketplace for resources, while simultaneously supporting a permissionless and free-to-use ecosystem.
This section describes the comprehensive design of our search engine through three key user interactions: \emph{search}, \emph{click}, and \emph{donate}.

\subsection{Decentralized Search}

Our search strategy follows the two-stage approach of \emph{retrieve-then-rank} common in modern search engines~\cite{nogueira2019passage,zhao2024dense,yates2021pretrained,paliwal2024cross}.
Accordingly, the retrieval of documents and the subsequent ranking of retrieved documents represent two distinct components.
Every user operates a search engine. Every user can answer queries.
Certain users receive donations from their fans, as they operate reliably and offer a high quality of service (in the form of retrieved documents or exchanged training data).

\subsubsection*{Generative Retrieval}

Our system adopts De-DSI~\cite{neague2024dsi}, a recently proposed design for decentralized semantic search.
The authors of De-DSI showed how small language models can be used to index a large number of documents whilst maintaining high retrieval accuracy.
Concretely, they proposed a system in which AI agents train local encoder-decoder transformer models on query-docid pairs and retrieve matching docids on request (\emph{docids} refer to \emph{document identifiers} and can represent, for instance, URLs). 
This paradigm is known as \emph{generative retrieval}, and has recently gained popularity in the field of information retrieval~\cite{tay2022transformer,li2024matching}.

The search is initiated by a user submitting a query (e.g., ``ubuntu download'').
The user then solicits matching documents from multiple agents in the network.
Once a query is received by an agent, it performs a \emph{beam search} on its model.
The model produces a fixed-sized set of (e.g., 5 or 10) docids $d_i$ that most likely match the query, along with associated logit scores $s_i$, forming beams in the form $(d_i, s_i)$. 
This result set, subsequently, is sent back to the querying user,
who performs a beam search on their own model as well.
The scores $s_i$ within each set of beams are normalized using the softmax function.
Further, all beams are aggregated into a single set of docid candidates, where the scores of equal docids are summed.
This procedure approximates the intractable task of fully enumerating the distribution of all possible docid generations. 
The underlying assumption is that, across an ensemble of similarly trained models, relevant documents will consistently appear among the top candidates from multiple models, even if their exact ranking varies slightly. 
Consequently, aggregating predictions from multiple models via this ensemble approach yields superior retrieval accuracy compared to relying on a single model.
Finally, the top-$k$ docids with the highest scores are selected from this initial retrieval phase to undergo further processing in the re-ranking stage.

\subsubsection*{Relevance Ranking}

After a set of matching documents has been retrieved,
the next stage involves refining the ranking of these documents.
This stage is crucial as it evaluates documents beyond their semantic similarity with the query.
Relevance ranking, moreover, has the ability to account for statistical features that indicate, for instance, the popularity of a document or the alignment with user-specific personalization signals.
To this end, our system adopts DART~\cite{gregoriadis2025decentralized}, a decentralized adoption of learning-to-rank~\cite{cao2007learning,liu2009learning}.
In DART, each peer trains a local model $\mathcal{R}: X \mapsto Y$ to predict relevance $Y$ for a set of documents $X$.
The individual documents are represented as feature vectors $x_i\in\mathbb{R}^f$,
which encode the relationship between the query and the document.
The output of the model maps each $x_i\in X$ to a score $y_i\in[0,1]$, indicating relevance relative to other documents in the result list.
The ranking model is very lightweight, allowing for fast inference even on low-end hardware.
Furthermore, due to the abstract nature of the feature vectors, DART only needs a few data points in order to generalize to unseen queries and unseen documents.
This allows relevance ranking to be executed \emph{offline}, avoiding networking latency.

\subsection{Click-based Feedback}

When a user selects a document from a search result list, we refer to it as a \emph{click}.
Clicks are central to our system, as they present the sole signal driving our search engine’s training.
We interpret clicks as the user deeming a specific document \emph{relevant} w.r.t. the query.
Also, we interpret every other document in the search result list as being deemed \emph{non-relevant} w.r.t. the query.
These \emph{clicklogs} serve as feedback to both the retrieval and the ranking model of our search engine, and are further disseminated in the network.
The models, therefore, continuously train on more and updated data, enabling them to adapt and improve over time.



\subsubsection*{Spam Prevention}\label{sec:design:spam}

Opening up the user's local model to the influence of strange peers exposes the model to low-quality training data as well as adversarial attacks.
The incentive structure of our system exacerbates this issue, as it encourages the cheap generation of fake data (i.e., spam), in exchange for utility tokens. 
It is, therefore, paramount to assess the quality of the training data offered by another user.

Data valuation is an active topic in the field of federated and decentralized data markets~\cite{hammoudeh2024training,lu2024data,lu2024daved,wang2020principled,ghorbani2019data}.
Its goal is to quantify the contribution of training examples to the model's performance.
Among existing techniques, Data Shapley~\cite{ghorbani2019data} remains the most widely adopted. 
According to Ghorbani et al., a principled and equitable approach to the valuation for any dataset $\mathcal{D}_i$ among $n$ total datasets must approximate

\begin{equation}\label{eq:shapley}
    \phi_i = \sum_{S \subseteq D \setminus \mathcal{D}_i} \frac{V(S \cup \mathcal{D}_i) - V(S)}{\binom{n-1}{|S|}}.
\end{equation}

Here, $V(D)$ denotes the performance of the model trained on dataset $D$.
In words, this formula measures the marginal contribution of dataset $\mathcal{D}_i$ by user $i$ as part of the full dataset $D$ on which the model is trained on. 
In practice, \Cref{eq:shapley} is approximated using methods like Monte Carlo~\cite{ghorbani2019data,kwon2021beta}.

Our system adopts Data Shapley to assign value to the datasets offered by other peers. 
Upon receiving $n$ datasets from $n$ distinct peers, 
the user performs a Monte Carlo simulation to estimate the marginal contribution of each dataset to the performance of their own model.
The user’s historical clicks serve as the ground truth for evaluation. 

\subsection{Decentralized Donations}
\label{sec:design:donations}
To support a self-sustaining ecosystem without fees or centralized infrastructure, our system introduces a decentralized donation mechanism. Rather than paying per-query, users can voluntarily donate tokens, such as Bitcoin or stablecoins, to support the search network. These donations are then distributed among peers based on their perceived utility, captured through a personalized and Sybil-tolerant\footnote{A Sybil attack is when one entity creates many fake identities to gain unfair influence in a system.} reputation mechanism known as MeritRank~\cite{nasrulin2022meritrank}.

Each node maintains a local ledger of feedback collected from its own search and training interactions. This feedback encompasses observed document relevance, ranking accuracy, and the value of shared training data. Over time, users build subjective views of the network, forming a weighted feedback graph. In this graph, edges represent feedback, and their weights reflect the perceived value of past contributions from peer nodes.

To determine how to distribute a donation, a user calculates utility scores based on their local feedback graph. MeritRank is a decentralized reputation algorithm that addresses the three-way tradeoff between Sybil resistance, generalizability, and trustlessness. It avoids reliance on a central scoring oracle or a fixed set of validators. Instead, it aggregates feedback locally while limiting the influence of malicious nodes through three decay functions:

\begin{itemize}
    \item Transitivity decay penalizes contributions from distant nodes in the feedback graph.
    \item Connectivity decay down-weights contributions from sparsely connected or weakly integrated nodes.
    \item Epoch decay discounts older feedback to maintain relevance.
\end{itemize}

The resulting scores form a subjective reputation vector over other nodes. While a simple donation scheme could distribute funds proportionally to these scores, we adopt a slightly more sophisticated mechanism to balance meritocracy with stochastic fairness. Specifically, we apply a \emph{weighted random sampling strategy}, ensuring that smaller or newer contributors maintain a non-zero probability of being rewarded, while high-ranking nodes remain strongly favored.

\begin{enumerate}
    \item \textbf{Weighting.} Each node $j$ is assigned a weight based on their local MeritRank score $R_j$ as follows:
    \[
    W_j = (R_j + \epsilon)^\alpha
    \]
    Here, $\epsilon > 0$ prevents zero-probability for low-ranked nodes, and $\alpha \geq 1$ controls the sharpness of merit preference.

    \item \textbf{Sampling.} Let $S$ denote the set of nodes with non-zero reputation scores $R_j$ in the donor's local view---i.e., those who have contributed utility either directly or indirectly through the feedback graph. From this set, the donor selects a subset of $N$ recipients using weighted random sampling without replacement, where each node $j$ is chosen with probability:
    \[
    P_j = \frac{W_j}{\sum_{k \in S} W_k}
    \]

    \item \textbf{Allocation.} Let $S' \subseteq S$ be the set of $N$ recipients selected during the sampling step. Given a total donation amount $D$, the funds are split among the sampled nodes proportionally:
    \[
    A_j = D \cdot \frac{W_j}{\sum_{k \in S'} W_k} \quad \text{for } j \in S'
    \]

    \item \textbf{Execution.} The donor then publishes a signed payout message listing all recipients and assigned values $A_j$. Transfers are settled on the underlying payment layer.
\end{enumerate}

This approach mirrors personalized funding models from systems like SourceCred~\cite{rennie2022toward} and Gitcoin~\cite{gitcoinGitcoinFund}, but removes reliance on global state or central coordination. By introducing a stochastic selection process, we ensure that while top contributors are rewarded more frequently, smaller or less connected nodes are not excluded outright. This mitigates long-term monopolization and encourages continued participation across the network.

Importantly, this donation mechanism remains Sybil-tolerant, meaning that malicious actors cannot amplify their rewards by creating fake identities. Since MeritRank uses decay to bound transitive influence and isolates poorly connected components, Sybil regions yield diminishing returns. Agents are rewarded for genuine utility, not presence.

Through this design, we establish a decentralized funding mechanism that enables long-term sustainability without requiring fees or trusted intermediaries. Even if only a small subset of users choose to donate, their donations are targeted with high precision, rewarding those agents that actually contributed value.

\section{Dataset}\label{sec:dataset}

In 2006, AOL released a dataset of query logs from users of their web search engine~\cite{pass2006picture}.
To this date, it remains the only publicly available dataset that combines user mappings, raw search queries, and the URL of the clicked document~\cite{macavaneysigir2021irds},
all of which are necessary for the experimental evaluation of our search engine.
More recently, MacAvaney et al.~\cite{macavaney2022reproducing} successfully restored the contents of the majority of documents.
By leveraging the Internet Archive\footnote{https://archive.org/},
their approach ensured that the recovered content approximated the state of the documents at the time of the user query.
Furthermore, they parsed the HTML, extracting title and body as plain text.

\begin{table*}[h]
    \centering
    \vspace{0.03in}
    \caption{Excerpt of Query Logs From Our Dataset}
    \begin{tabularx}{\textwidth}{llllll}
    \toprule
    User ID & Time & Query & Doc ID & Candidate Doc IDs \\
    \midrule
    3613173 & 2006-03-01 00:01:04 &                 batman signal images &  b093fab50ffa &  [8bf002ac9afd, 07098e6e8908, a411c4f4616e, f60... \\
1270972 & 2006-03-01 00:01:10 &                         head hunters &  cba56bcc7234 & [718c5195a5d4, 9cdcd810b216, a7fb5708b128, a9e... \\
811283 & 2006-03-01 00:01:16   &     new jersey inside water parks &  44792c7f06a7  & [4c080044d7d0, c72c4ea94b4d, 95168237c7ad, a04... \\
2648672 & 2006-03-01 00:01:17 &                           jamie farr &  545bf737bf10 &  [7dafa223564f, fed9f2a111eb, 51b70ee252d5, dda... \\
605089 & 2006-03-01 00:01:18 &  dachshunds for sale in conroe texas &  ccc0c7d8dce5 &  [a93366fdaa88, 5162ff424e92, bc596fea9ed1, 0f5...
    \end{tabularx}
    \label{tab:dataset:queries}
\end{table*}

In addition, the training of our ranking model requires the non-clicked documents given a query, i.e., the list of displayed search results given a search query.
This data is missing from the dataset and cannot be restored to its original form.
To address this limitation, we generate candidate document sets based on the methodology proposed by Guo et al.~\cite{guo2021aol4ps}.
Specifically, given each query, we rank the entire set of documents based on the BM25 score of the document titles.
BM25 is a standard keyword-based ranking function used in information retrieval~\cite{robertson1995okapi,robertson2009probabilistic}.
We then apply a window of size 10 on the ranked list with the clicked document in the center.
That is, if the clicked document is ranked at position 20 in the BM25-ordered list, we consider the documents from positions 15--24 the result candidates for the given query.
If the position of the clicked document is close to the bounds of the list, we extend the selection accordingly so that we always return 10 candidate documents.
In \Cref{tab:dataset:queries}, we show an excerpt of our datasets, consisting of query logs, the clicked document (``Doc ID''), as well as the list of candidates that we compiled. 

Finally, we apply the following filters.
We removed query logs that were not associated with a clicked document (as the original dataset also contained queries with no further user feedback).
Furthermore, we removed duplicate (User ID, Query) pairs, whereby we only kept the most recent record based on its timestamp.

\section{Experiments}\label{sec:exp}

We established that a decentralized search engine needs to deliver high-quality results and, at the same time, be resistant to attacks, in order to be able to compete with centralized alternatives.
In this section, we demonstrate how our system meets both requirements throughout three experiments based on the dataset presented in \Cref{sec:dataset}.
The first experiment demonstrates the accuracy of document retrieval under non-adversarial conditions, as well as its scalability w.r.t the number of documents.
In the second experiment, we demonstrate the effectiveness of the Data Shapley algorithm in mitigating the impact of spam.
Our final experiment measures the resilience of our financial architecture against Sybil attacks.
We release our dataset and experimental code as open-source\footnote{https://github.com/mg98/swarmsearch}.

\subsection{Retrieval Accuracy}

Every agent operates a transformer model for retrieval.
The retrieval accuracy of a single model is constrained by its capacity to recall a large number of documents.
Collectively, however, they form a network of retrieval engines that can scale to any number of documents.

In this experiment, we show the retrieval accuracy of a single retrieval model given a corpus of 100, 1000, and 5000 documents.
As a basis for this experiment, we fine-tune a T5-base model~\cite{ni2021sentence}, which is a popular pre-trained language model.
With 220 million parameters, we consider its size reasonable for deployment on commodity hardware.
We randomly sample 100, 1000, and 5000 documents from our dataset.
For each document, we collect all associated queries in chronological order and partition them into training and test sets using a 90:10 split.
To support this split, we only included documents with at least 10 associated queries.
Importantly, we do not remove query duplicates across the splits.
That is, we allow for the same query-docid pair to appear in both the training and test set.
This is atypical when evaluating the quality of machine learning models.
However, we intend to capture realistic usage patterns and real-world settings.
That is, if there is a high likelihood for a query in the test set, this likelihood should also be reflected for its inclusion in the training set (i.e., the data it has already seen and trained on).
Finally, we fine-tune a T5-base model on the collection of training splits of all documents in the current experiment (i.e., the sample of either 100, 1000, or 5000).
The training set itself is shuffled and deduplicated, to avoid overfitting.
Lastly, we evaluate on the collection of test splits of all documents.
To this end, we only consider matches on the top-1 beam.

\begin{figure}[t]
\centering
\input{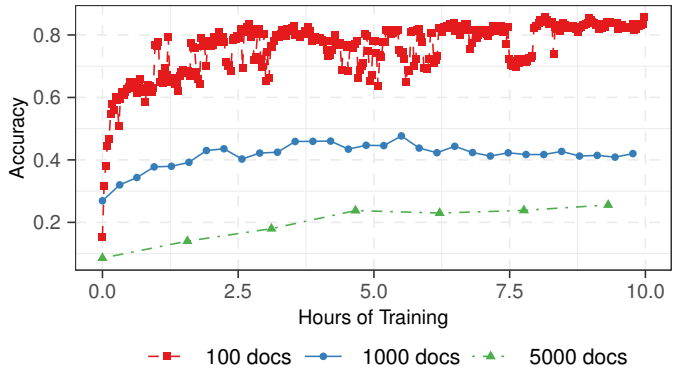}
\caption{Top-1 document retrieval accuracy of a single model evaluated on a test set of queries. Each data point represents one epoch of training. The x-axis is truncated where convergence was observed across all lines.}
\label{fig:dsi_exp}
\end{figure}

In \Cref{fig:dsi_exp}, we show the accuracy of this experiment as it converges over multiple epochs.
As the data volume differs between experiments on 100, 1000, and 5000 documents, we normalize the x-axis by \emph{hours of training} rather than by epochs.
We ran the experiments for 50 hours, but truncated the plot at the 10 hour-mark, where we observed convergence.
As the figure reveals, with 100 documents, the retrieval accuracy is nearly perfect, considering that users often choose different documents for the same query, and therefore an accuracy close to \qty{100}{\%} is unattainable.
With larger corpora of documents, we can see the model's performance degrading--at least in top-1 retrieval.
This does not inherently pose a problem, as the correct document may still appear in a top-5 or top-10 beam search before its final ranking position is determined by the querying user's ranking model.

\subsection{Spam Prevention}

With data sharing being lucrative, it creates an incentive for users to generate and disseminate inauthentic data for profit.
Accordingly, we introduced spam prevention measures in \Cref{sec:design:spam}.
In this experiment, we demonstrate their effectiveness, exemplified by the accuracy of the ranking model, as measured by \emph{Mean Reciprocal Rank (MRR)}.
MRR is a standard metric for assessing ranking quality~\cite{voorhees1999trec,xu2007adarank}.
Given a ranked list of search results, we compute the reciprocal rank $1/r$, where $r$ is the position of the item the user \emph{clicks}.
Averaging this measure across multiple queries yields the MRR.
Recall that our dataset guarantees exactly 10 search results containing 1 relevant item for each query.
That is, the lowest possible MRR in this setting is $\frac{1}{10}$, whereas a random ranking strategy would yield an expected MRR of $\frac{1}{5.5}\approx0.18$.

For our experiment, we simulate users connected to 10 randomly sampled users from our dataset; in the following referred to as \emph{neighbors}.
Preliminary experiments revealed converged performance with this number of neighbors.
We only consider users with at least three clicklogs.
The clicklogs of a user constitute a user's \emph{personal dataset}.
First, we partition each user’s personal dataset into training, validation, and test sets using a 1:1:1 ratio.
This split ratio was chosen to maximize the number of users meeting the minimum data threshold.

\subsubsection*{Local-only}
On this basis, we train and evaluate the ranking model with ``local-only'' data.
We observe an MRR of 0.38 when users train exclusively on local training data.

\subsubsection*{Data Sharing}
Ranking accuracy with local-only data is low, as training data is scarce.
In the next step, we fuse the personal datasets of the user's neighbors with the user's local training data.
That is, we augment the training split with the training data from all 10 neighbors, while retaining the original validation and test sets.
On average, in this setting, users yield an MRR of 0.67.
That is, users on average find the relevant item (as indicated by their clicks) ranked at position 1 or 2.

\subsubsection*{Adversarial Data Sharing}

\begin{figure}[ht]
    \centering
    \input{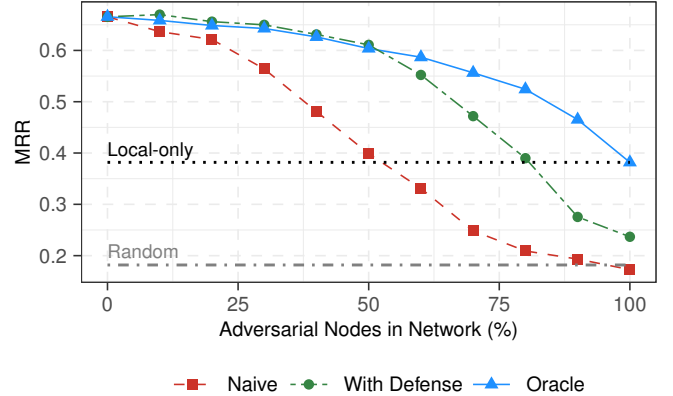}
    \vspace{-3em}
    \caption{Ranking accuracy under increasing share of malicious agents performing poisoning attacks.}
    \label{fig:poisoning}
\end{figure}

Next, we analyze the effects of adversarial nodes on the ranking performance.
To this end, we apply label flipping to $n\in[0,10]$ out of the 10 datasets presented to a user,
essentially simulating a poisoning attack~\cite{tolpegin2020data,paudice2019label}.
In \Cref{fig:poisoning}, we show the performance degradation under an increasing share of adversarial nodes $n/10$.
We show three modes of operation:
\begin{itemize}
    \item \textbf{Naive:} Users assume all data is valuable and incorporate every dataset into training without scrutiny.
    \item \textbf{With Defense:} Users approximate the Shapley value $\phi_i$ of each dataset using a Monte Carlo simulation, with a permutation sample size of 100, and only incorporate datasets where $\phi_i>0$.
    \item \textbf{Oracle:} For reference, we report an oracle baseline that excludes any poisoned datasets.
\end{itemize}

As expected, as soon as we introduce adversarial nodes ($n>0$), the MRR begins to decline.
At up to \qty{60}{\%} with the naive strategy and \qty{80}{\%} with defense,
the ranking performance still exceeds performance under local-only training.
Notably, our defense mechanism tolerates up to \qty{50}{\%} adversarial nodes without any performance degradation; 
beyond this threshold, performance declines more rapidly.
As the amount of ``beneficial'' data diminishes, we would expect the ranking accuracy converging to local-only performance.
However, at \qty{90}{\%}, it drops below.
We attribute this to the Shapley-value approximation error, 
which we expect could be reduced by employing larger sample sizes~\cite{maleki2013bounding,wu2023variance} or more efficient algorithms~\cite{kwon2021beta,jia2019efficient,wang2023data}.

\subsection{Resilience Against Sybil Attacks}

To evaluate the robustness of our decentralized donation mechanism, we simulate a series of Sybil attack scenarios and measure the percentage of total donations captured by malicious nodes as their proportion of the network increases. Our aim is to test the mechanism's ability to uphold meritocratic allocation while resisting manipulation.

We simulate a search network of 1000 user nodes, modeled after behavioral patterns in our dataset, where a varying fraction (0--100\,\%) is controlled by a Sybil adversary. Honest nodes receive reputation scores sampled from our real-world dataset, containing anonymized records of user search activity, including queries, document candidates, and clicks. Each unique ``User ID'' is treated as a user node, and reputation scores are computed based on total click activity under the assumption that consistent interaction with search results correlates with meaningful participation in the system. Click counts are normalized to the range $[0, 1]$, yielding a continuous reputation distribution that reflects relative contribution levels across the user population.
To simulate donations, each round begins with a designated donor node allocating a fixed amount across a subset of peers using the weighted lottery-based mechanism described in \Cref{sec:design:donations}. Sybil nodes are assigned low or manipulated reputation scores depending on the scenario, either by sampling from the lower tail of the honest distribution or by injecting dense feedback among themselves. For each configuration, we simulate multiple donation rounds and report the average donation share received by Sybil nodes.

\begin{figure}[t]
\centering
\input{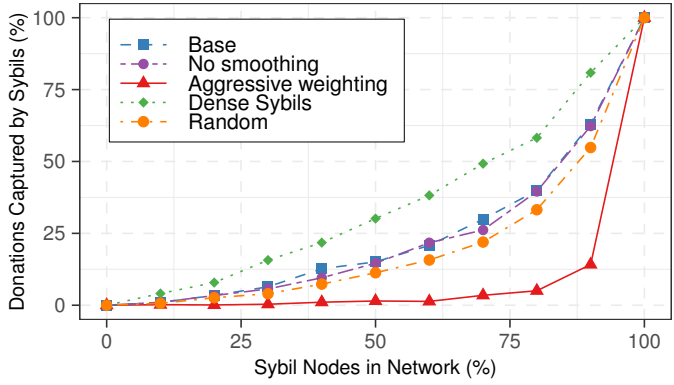}
\caption{Donation captured by Sybil nodes under different attack and configuration scenarios.}
\label{fig:sybil}
\end{figure}

\Cref{fig:sybil} presents the results across five scenarios:

\begin{itemize}
    \item \textbf{Base:} Honest reputations are sampled from our dataset, with $\alpha = 0.5$ and $\epsilon = 0.01$. Sybils have uniformly low utility scores. This represents a balanced system with a mild preference for high-reputation nodes and a small smoothing factor, allowing low-ranked nodes a chance.
    \item \textbf{No smoothing:} The same setup as the base case, but  $\epsilon=0$. This eliminates stochastic inclusion of low-reputation nodes. As expected, Sybil capture drops sharply---demonstrating the protective role of hard cutoffs---but also implies reduced robustness to newcomers.
    \item \textbf{Aggressive weighting:} Here, $\alpha = 1.5$ increases the reward skew toward top-ranked nodes. Honest nodes dominate allocation at low Sybil rates. 
    Under higher Sybil concentrations, as attacker nodes inflate utility scores within isolated clusters, the system becomes vulnerable.
    \item \textbf{Dense Sybils:} A more realistic attack model in which Sybils assign feedback to each other and maintain moderate reputations. This scenario illustrates that when attacker nodes escape the lowest ranks, capture rates rise significantly, underscoring the importance of connectivity and decay in MeritRank to limit transitive influence.
    \item \textbf{Random:} Honest reputations are drawn from a uniform distribution rather than empirical data. This scenario serves as a sanity check and exhibits similar trends to the base case, albeit with slightly higher variance in Sybil capture.
\end{itemize}

Together, these experiments highlight that while the lottery-based mechanism remains resilient under various configurations, parameters such as $\epsilon$ and $\alpha$ introduce trade-offs between openness, fairness, and robustness. In particular, mechanisms that rely solely on reputation without smoothing can fully exclude attackers but risk excluding legitimate newcomers, while overly flat weighting can make the system vulnerable to low-effort Sybil strategies.

\section{Conclusion}\label{sec:conclusion}

Decentralization of search has been an open problem for more than 25 years.
We believe that our financial architecture provides the previously missing foundation that enables the adoption of effective search algorithms, especially those leveraging artificial intelligence.
With \textsc{SwarmSearch}, we demonstrated how state-of-the-art search can be supported by an agent-to-agent economy without centralized coordination.
Through extensive experiments, we also validated its robustness to spam and Sybil attacks, showing the viability of a permissionless and self-sustaining search infrastructure.

In future work, we want to focus on the scalability of our system alongside privacy-preserving techniques.
We are working towards the deployment of our search engine within the peer-to-peer file-sharing software Tribler.
Through our continuous contributions, we hope that we can soon achieve a decentralized search engine able to rival centralized alternatives.

\bibliographystyle{IEEEtran}
\bibliography{references}

\begin{thebibliography}{10}
\providecommand{\url}[1]{#1}
\csname url@samestyle\endcsname
\providecommand{\newblock}{\relax}
\providecommand{\bibinfo}[2]{#2}
\providecommand{\BIBentrySTDinterwordspacing}{\spaceskip=0pt\relax}
\providecommand{\BIBentryALTinterwordstretchfactor}{4}
\providecommand{\BIBentryALTinterwordspacing}{\spaceskip=\fontdimen2\font plus
\BIBentryALTinterwordstretchfactor\fontdimen3\font minus \fontdimen4\font\relax}
\providecommand{\BIBforeignlanguage}[2]{{%
\expandafter\ifx\csname l@#1\endcsname\relax
\typeout{** WARNING: IEEEtran.bst: No hyphenation pattern has been}%
\typeout{** loaded for the language `#1'. Using the pattern for}%
\typeout{** the default language instead.}%
\else
\language=\csname l@#1\endcsname
\fi
#2}}
\providecommand{\BIBdecl}{\relax}
\BIBdecl

\bibitem{usdoj_google_2023}
\BIBentryALTinterwordspacing
{U.S. Department of Justice}, ``Justice department sues google for monopolizing digital advertising technologies,'' 2023, accessed: 2024-09-10. [Online]. Available: \url{https://www.justice.gov/opa/pr/justice-department-sues-google-monopolizing-digital-advertising-technologies}
\BIBentrySTDinterwordspacing

\bibitem{epstein2016new}
R.~Epstein, ``The new censorship,'' \emph{US News}, vol.~22, 2016.

\bibitem{solon2016google}
O.~Solon and S.~Levin, ``How google’s search algorithm spreads false information with a rightwing bias,'' \emph{The Guardian}, vol.~16, 2016.

\bibitem{stjernfelt2020facebook}
F.~Stjernfelt, A.~M. Lauritzen, F.~Stjernfelt, and A.~M. Lauritzen, ``Facebook and google as offices of censorship,'' \emph{Your post has been removed: tech giants and freedom of speech}, pp. 139--172, 2020.

\bibitem{hinds2020wouldn}
J.~Hinds, E.~J. Williams, and A.~N. Joinson, ``“it wouldn't happen to me”: Privacy concerns and perspectives following the cambridge analytica scandal,'' \emph{International Journal of Human-Computer Studies}, vol. 143, p. 102498, 2020.

\bibitem{eu2024tiktok}
E.~Commission, ``Commission opens formal proceedings against tiktok on election risks under the digital services act,'' Press release IP/24/6487. Available from \url{https://ec.europa.eu/commission/presscorner/detail/en/ip_24_6487}, December 2024.

\bibitem{chawathe2003making}
Y.~Chawathe, S.~Ratnasamy, L.~Breslau, N.~Lanham, and S.~Shenker, ``Making gnutella-like p2p systems scalable,'' in \emph{Proceedings of the 2003 conference on Applications, technologies, architectures, and protocols for computer communications}, 2003, pp. 407--418.

\bibitem{keizer2024survey}
N.~Keizer, O.~Ascigil, M.~Kr{\'o}l, D.~Kutscher, and G.~Pavlou, ``A survey on content retrieval on the decentralised web,'' \emph{ACM Computing Surveys}, vol.~56, no.~8, pp. 1--39, 2024.

\bibitem{khudhur2019siva}
N.~Khudhur and S.~Fujita, ``Siva-the ipfs search engine,'' in \emph{2019 Seventh International Symposium on Computing and Networking (CANDAR)}.\hskip 1em plus 0.5em minus 0.4em\relax IEEE, 2019, pp. 150--156.

\bibitem{li2021bringing}
M.~Li, J.~Zhu, T.~Zhang, C.~Tan, Y.~Xia, S.~Angel, and H.~Chen, ``Bringing decentralized search to decentralized services,'' in \emph{15th $\{$USENIX$\}$ Symposium on Operating Systems Design and Implementation ($\{$OSDI$\}$ 21)}, 2021, pp. 331--347.

\bibitem{loo2005case}
B.~T. Loo, R.~Huebsch, I.~Stoica, and J.~M. Hellerstein, ``The case for a hybrid p2p search infrastructure,'' in \emph{Peer-to-Peer Systems III: Third International Workshop, IPTPS 2004, La Jolla, CA, USA, February 26-27, 2004, Revised Selected Papers 3}.\hskip 1em plus 0.5em minus 0.4em\relax Springer, 2005, pp. 141--150.

\bibitem{wang2020keyword}
F.~Wang and Y.~Wu, ``Keyword search technology in content addressable storage system,'' in \emph{2020 IEEE 22nd Intl. Conference on High Performance Computing and Communications; IEEE 18th Intl. Conference on Smart City; IEEE 6th Intl. Conference on Data Science and Systems (HPCC/SmartCity/DSS)}.\hskip 1em plus 0.5em minus 0.4em\relax IEEE, 2020, pp. 728--735.

\bibitem{lu2002panache}
T.~Lu, S.~Sinha, and A.~Sudan, ``Panache: A scalable distributed index for keyword search,'' Citeseer, Tech. Rep., 2002.

\bibitem{herrmann2014description}
M.~Herrmann, K.-C. Ning, C.~Diaz, and B.~Preneel, ``Description of the yacy distributed web search engine,'' \emph{Technicla report. KU Leuven ESAT/COSIC, iMinds}, 2014.

\bibitem{onal2018neural}
K.~D. Onal, Y.~Zhang, I.~S. Altingovde, M.~M. Rahman, P.~Karagoz, A.~Braylan, B.~Dang, H.-L. Chang, H.~Kim, Q.~McNamara \emph{et~al.}, ``Neural information retrieval: At the end of the early years,'' \emph{Information Retrieval Journal}, vol.~21, pp. 111--182, 2018.

\bibitem{archiveBumpRoad}
F.~Emans, ``{B}ump in the road --- web.archive.org,'' \url{https://web.archive.org/web/20240422190159/https://blog.ipfs-search.com/bump-in-the-road/}, 2023, [Accessed 10-09-2024].

\bibitem{poort2014baywatch}
J.~Poort, J.~Leenheer, J.~van~der Ham, and C.~Dumitru, ``Baywatch: Two approaches to measure the effects of blocking access to the pirate bay,'' \emph{Telecommunications Policy}, vol.~38, no.~4, pp. 383--392, 2014.

\bibitem{presearchPresearch}
``{P}research --- presearch.com,'' \url{https://presearch.com/}, [Accessed 16-08-2024].

\bibitem{wei2015motivating}
X.~Wei, W.~Chen, and K.~Zhu, ``Motivating user contributions in online knowledge communities: virtual rewards and reputation,'' in \emph{2015 48th Hawaii Intl. Conf. on System Sciences}.\hskip 1em plus 0.5em minus 0.4em\relax IEEE, 2015, pp. 3760--3769.

\bibitem{goes2016incentive}
P.~B. Goes, C.~Guo, and M.~Lin, ``Do incentive hierarchies induce user effort? evidence from an online knowledge exchange,'' \emph{Information Systems Research}, vol.~27, no.~3, pp. 497--516, 2016.

\bibitem{qiao2021mitigating}
D.~Qiao, S.-Y. Lee, A.~B. Whinston, and Q.~Wei, ``Mitigating the adverse effect of monetary incentives on voluntary contributions online,'' \emph{Journal of Management Information Systems}, vol.~38, no.~1, pp. 82--107, 2021.

\bibitem{ecbMicropayments}
E.~C. Bank, \emph{A big future for small payments? -- Micropayments and their impact on the payment ecosystem}.\hskip 1em plus 0.5em minus 0.4em\relax European Central Bank, 2023.

\bibitem{solaiman2024internet}
E.~Solaiman and J.~Robins, ``The internet of value: Integrating blockchain and lightning network micropayments for knowledge markets,'' \emph{arXiv preprint arXiv:2412.19384}, 2024.

\bibitem{ngoc2006maay}
F.~D. Ngoc, J.~Keller, and G.~Simon, ``Maay: a decentralized personalized search system,'' in \emph{International Symposium on Applications and the Internet (SAINT'06)}.\hskip 1em plus 0.5em minus 0.4em\relax IEEE, 2006, pp. 8--pp.

\bibitem{pouwelse2008tribler}
J.~A. Pouwelse, P.~Garbacki, J.~Wang, A.~Bakker, J.~Yang, A.~Iosup, D.~H. Epema, M.~Reinders, M.~R. Van~Steen, and H.~J. Sips, ``Tribler: a social-based peer-to-peer system,'' \emph{Concurrency and computation: Practice and experience}, vol.~20, no.~2, pp. 127--138, 2008.

\bibitem{thegraph}
\BIBentryALTinterwordspacing
B.~Ramirez, ``The graph network in depth - part 1,'' Oct 2019. [Online]. Available: \url{https://thegraph.com/blog/the-graph-network-in-depth-part-1}
\BIBentrySTDinterwordspacing

\bibitem{gregoriadis2025decentralized}
M.~Gregoriadis, Q.~Stokkink, and J.~Pouwelse, ``Decentralized adaptive ranking using transformers,'' in \emph{Proceedings of the 5th Workshop on Machine Learning and Systems}, 2025, pp. 12--18.

\bibitem{neague2024dsi}
P.~Neague, M.~Gregoriadis, and J.~Pouwelse, ``De-dsi: Decentralised differentiable search index,'' in \emph{Proceedings of the 4th Workshop on Machine Learning and Systems}, 2024, pp. 134--143.

\bibitem{neague2025semantica}
P.~Neague, Q.~Stokkink, N.~Goel, and J.~Pouwelse, ``Semantica: Decentralized search using a llm-guided semantic tree overlay,'' \emph{arXiv preprint arXiv:2502.10151}, 2025.

\bibitem{ripeanu2001peer}
M.~Ripeanu, ``Peer-to-peer architecture case study: Gnutella network,'' in \emph{Proceedings first international conference on peer-to-peer computing}.\hskip 1em plus 0.5em minus 0.4em\relax IEEE, 2001, pp. 99--100.

\bibitem{adar2000free}
\BIBentryALTinterwordspacing
E.~Adar and B.~A. Huberman, ``Free riding on gnutella,'' \emph{First Monday}, vol.~5, no.~10, Oct. 2000. [Online]. Available: \url{https://firstmonday.org/ojs/index.php/fm/article/view/792}
\BIBentrySTDinterwordspacing

\bibitem{Alphabet2025SEC}
\BIBentryALTinterwordspacing
{U.S. Securities and Exchange Commission}, ``{Form 10-K: Annual report for Alphabet Inc.}'' 2024, {Accessed: 2025-05-04}. [Online]. Available: \url{https://www.sec.gov/Archives/edgar/data/1652044/000165204425000014/goog-20241231.htm}
\BIBentrySTDinterwordspacing

\bibitem{p2pfoundationFreemiumBusiness}
P.~Froberg, ``{F}reemium, a business model for {P}2{P} | {P}2{P} {F}oundation --- blog.p2pfoundation.net,'' \url{https://blog.p2pfoundation.net/freemium-a-business-model-for-p2p/2008/12/08}, 2008, [Accessed 04-05-2025].

\bibitem{rietveld2018creating}
J.~Rietveld, ``Creating and capturing value from freemium business models: A demand-side perspective,'' \emph{Strategic Entrepreneurship Journal}, vol.~12, no.~2, pp. 171--193, 2018.

\bibitem{gassend2001dinx}
B.~Gassend, T.~M. Gil, and B.~Song, ``Dinx: A decentralized search engine,'' 2001.

\bibitem{zhu2020keyword}
L.~Zhu, C.~Xiao, and X.~Gong, ``Keyword search in decentralized storage systems,'' \emph{Electronics}, vol.~9, no.~12, p. 2041, 2020.

\bibitem{keizer2023ditto}
N.~V. Keizer, O.~Ascigil, M.~Kr{\'o}l, and G.~Pavlou, ``Ditto: Towards decentralised similarity search for web3 services,'' in \emph{2023 IEEE International Conference on Decentralized Applications and Infrastructures (DAPPS)}.\hskip 1em plus 0.5em minus 0.4em\relax IEEE, 2023, pp. 66--75.

\bibitem{fujita2021similarity}
S.~Fujita, ``Similarity search in interplanetary file system with the aid of locality sensitive hash,'' \emph{IEICE TRANSACTIONS on Information and Systems}, vol. 104, no.~10, pp. 1616--1623, 2021.

\bibitem{gold2023g}
A.~Gold and J.~Pouwelse, ``G-rank: Unsupervised continuous learn-to-rank for edge devices in a p2p network,'' \emph{arXiv preprint arXiv:2301.12530}, 2023.

\bibitem{presearchWhatNode}
``{W}hat is a {N}ode? | {P}research {D}ocs --- docs.presearch.io,'' \url{https://docs.presearch.io/nodes/overview}, [Accessed 04-05-2025].

\bibitem{nogueira2019passage}
R.~Nogueira and K.~Cho, ``Passage re-ranking with bert,'' \emph{arXiv preprint arXiv:1901.04085}, 2019.

\bibitem{zhao2024dense}
W.~X. Zhao, J.~Liu, R.~Ren, and J.-R. Wen, ``Dense text retrieval based on pretrained language models: A survey,'' \emph{ACM Transactions on Information Systems}, vol.~42, no.~4, pp. 1--60, 2024.

\bibitem{yates2021pretrained}
A.~Yates, R.~Nogueira, and J.~Lin, ``Pretrained transformers for text ranking: Bert and beyond,'' in \emph{Proceedings of the 14th ACM International Conference on web search and data mining}, 2021, pp. 1154--1156.

\bibitem{paliwal2024cross}
B.~Paliwal, D.~Saini, M.~Dhawan, S.~Asokan, N.~Natarajan, S.~Aggarwal, P.~Malhotra, J.~Jiao, and M.~Varma, ``Cross-jem: Accurate and efficient cross-encoders for short-text ranking tasks,'' \emph{arXiv preprint arXiv:2409.09795}, 2024.

\bibitem{tay2022transformer}
Y.~Tay, V.~Tran, M.~Dehghani, J.~Ni, D.~Bahri, H.~Mehta, Z.~Qin, K.~Hui, Z.~Zhao, J.~Gupta \emph{et~al.}, ``Transformer memory as a differentiable search index,'' \emph{Advances in Neural Information Processing Systems}, vol.~35, pp. 21\,831--21\,843, 2022.

\bibitem{li2024matching}
X.~Li, J.~Jin, Y.~Zhou, Y.~Zhang, P.~Zhang, Y.~Zhu, and Z.~Dou, ``From matching to generation: A survey on generative information retrieval,'' \emph{ACM Transactions on Information Systems}, 2024.

\bibitem{cao2007learning}
Z.~Cao, T.~Qin, T.-Y. Liu, M.-F. Tsai, and H.~Li, ``Learning to rank: from pairwise approach to listwise approach,'' in \emph{Proceedings of the 24th international conference on Machine learning}, 2007, pp. 129--136.

\bibitem{liu2009learning}
T.-Y. Liu \emph{et~al.}, ``Learning to rank for information retrieval,'' \emph{Foundations and Trends{\textregistered} in Information Retrieval}, vol.~3, no.~3, pp. 225--331, 2009.

\bibitem{hammoudeh2024training}
Z.~Hammoudeh and D.~Lowd, ``Training data influence analysis and estimation: A survey,'' \emph{Machine Learning}, vol. 113, no.~5, pp. 2351--2403, 2024.

\bibitem{lu2024data}
C.~Lu, M.~M. Amiri, and R.~Raskar, ``Data measurements for decentralized data markets,'' \emph{arXiv preprint arXiv:2406.04257}, 2024.

\bibitem{lu2024daved}
C.~Lu, B.~Huang, S.~P. Karimireddy, P.~Vepakomma, M.~Jordan, and R.~Raskar, ``Data acquisition via experimental design for data markets,'' \emph{Advances in Neural Information Processing Systems}, vol.~37, pp. 118\,086--118\,118, 2024.

\bibitem{wang2020principled}
T.~Wang, J.~Rausch, C.~Zhang, R.~Jia, and D.~Song, ``A principled approach to data valuation for federated learning,'' \emph{Federated Learning: Privacy and Incentive}, pp. 153--167, 2020.

\bibitem{ghorbani2019data}
A.~Ghorbani and J.~Zou, ``Data shapley: Equitable valuation of data for machine learning,'' in \emph{International conference on machine learning}.\hskip 1em plus 0.5em minus 0.4em\relax PMLR, 2019, pp. 2242--2251.

\bibitem{kwon2021beta}
Y.~Kwon and J.~Zou, ``Beta shapley: a unified and noise-reduced data valuation framework for machine learning,'' \emph{arXiv preprint arXiv:2110.14049}, 2021.

\bibitem{nasrulin2022meritrank}
B.~Nasrulin, G.~Ishmaev, and J.~Pouwelse, ``Meritrank: Sybil tolerant reputation for merit-based tokenomics,'' in \emph{2022 4th Conference on Blockchain Research \& Applications for Innovative Networks and Services (BRAINS)}.\hskip 1em plus 0.5em minus 0.4em\relax IEEE, 2022, pp. 95--102.

\bibitem{rennie2022toward}
E.~Rennie, M.~Zargham, J.~Tan, L.~Miller, J.~Abbott, K.~Nabben, and P.~De~Filippi, ``Toward a participatory digital ethnography of blockchain governance,'' \emph{Qualitative Inquiry}, vol.~28, no.~7, pp. 837--847, 2022.

\bibitem{gitcoinGitcoinFund}
``{G}itcoin | {F}und {W}hat {M}atters {T}o {Y}our {C}ommunity --- gitcoin.co,'' \url{https://www.gitcoin.co/}, [Accessed 08-05-2025].

\bibitem{pass2006picture}
G.~Pass, A.~Chowdhury, and C.~Torgeson, ``A picture of search,'' in \emph{Proceedings of the 1st international conference on Scalable information systems}, 2006, pp. 1--es.

\bibitem{macavaneysigir2021irds}
S.~MacAvaney, A.~Yates, S.~Feldman, D.~Downey, A.~Cohan, and N.~Goharian, ``Simplified data wrangling with ir\_datasets,'' in \emph{SIGIR}, 2021.

\bibitem{macavaney2022reproducing}
S.~MacAvaney, C.~Macdonald, and I.~Ounis, ``Reproducing personalised session search over the aol query log,'' in \emph{European Conference on Information Retrieval}.\hskip 1em plus 0.5em minus 0.4em\relax Springer, 2022, pp. 627--640.

\bibitem{guo2021aol4ps}
Q.~Guo, W.~Chen, and H.~Wan, ``Aol4ps: a large-scale data set for personalized search,'' \emph{Data Intelligence}, vol.~3, no.~4, pp. 548--567, 2021.

\bibitem{robertson1995okapi}
S.~E. Robertson, S.~Walker, S.~Jones, M.~M. Hancock-Beaulieu, M.~Gatford \emph{et~al.}, ``Okapi at trec-3,'' \emph{Nist Special Publication Sp}, vol. 109, p. 109, 1995.

\bibitem{robertson2009probabilistic}
S.~Robertson, H.~Zaragoza \emph{et~al.}, ``The probabilistic relevance framework: Bm25 and beyond,'' \emph{Foundations and Trends{\textregistered} in Information Retrieval}, vol.~3, no.~4, pp. 333--389, 2009.

\bibitem{ni2021sentence}
J.~Ni, G.~H. Abrego, N.~Constant, J.~Ma, K.~B. Hall, D.~Cer, and Y.~Yang, ``Sentence-t5: Scalable sentence encoders from pre-trained text-to-text models,'' \emph{arXiv preprint arXiv:2108.08877}, 2021.

\bibitem{voorhees1999trec}
E.~M. Voorhees, ``The trec-8 question answering track report.'' in \emph{Trec}, vol.~99, 1999, pp. 77--82.

\bibitem{xu2007adarank}
J.~Xu and H.~Li, ``Adarank: a boosting algorithm for information retrieval,'' in \emph{Proceedings of the 30th annual intl. ACM SIGIR conf. on Research and development in information retrieval}, 2007, pp. 391--398.

\bibitem{tolpegin2020data}
V.~Tolpegin, S.~Truex, M.~E. Gursoy, and L.~Liu, ``Data poisoning attacks against federated learning systems,'' in \emph{Computer security--ESORICs 2020: 25th European symposium on research in computer security, ESORICs 2020, guildford, UK, September 14--18, 2020, proceedings, part i 25}.\hskip 1em plus 0.5em minus 0.4em\relax Springer, 2020, pp. 480--501.

\bibitem{paudice2019label}
A.~Paudice, L.~Mu{\~n}oz-Gonz{\'a}lez, and E.~C. Lupu, ``Label sanitization against label flipping poisoning attacks,'' in \emph{ECML PKDD 2018 Workshops: Nemesis 2018, UrbReas 2018, SoGood 2018, IWAISe 2018, and Green Data Mining 2018, Dublin, Ireland, September 10-14, 2018, Proceedings 18}.\hskip 1em plus 0.5em minus 0.4em\relax Springer, 2019, pp. 5--15.

\bibitem{maleki2013bounding}
S.~Maleki, L.~Tran-Thanh, G.~Hines, T.~Rahwan, and A.~Rogers, ``Bounding the estimation error of sampling-based shapley value approximation with/without stratifying,'' \emph{CoRR, abs/1306.4265}, vol.~2, no.~1, 2013.

\bibitem{wu2023variance}
M.~Wu, R.~Jia, C.~Lin, W.~Huang, and X.~Chang, ``Variance reduced shapley value estimation for trustworthy data valuation,'' \emph{Computers \& Operations Research}, vol. 159, p. 106305, 2023.

\bibitem{jia2019efficient}
R.~Jia, D.~Dao, B.~Wang, F.~A. Hubis, N.~M. Gurel, B.~Li, C.~Zhang, C.~J. Spanos, and D.~Song, ``Efficient task-specific data valuation for nearest neighbor algorithms,'' \emph{arXiv preprint arXiv:1908.08619}, 2019.

\bibitem{wang2023data}
J.~T. Wang and R.~Jia, ``Data banzhaf: A robust data valuation framework for machine learning,'' in \emph{International Conference on Artificial Intelligence and Statistics}.\hskip 1em plus 0.5em minus 0.4em\relax PMLR, 2023, pp. 6388--6421.

\end{thebibliography}


\end{document}